\documentclass[twocolumn,showpacs,preprintnumbers,amsmath,amssymb,superscriptaddress]{revtex4}

\usepackage{graphicx}
\usepackage{dcolumn}
\usepackage{bm}
\usepackage{color}

\begin{document}

\title{Accessing surface Brillouin zone and band structure of picene single crystals}

\author{Qian Xin}\email{xin@restaff.chiba-u.jp}
\author{Steffen Duhm}
\author{Fabio Bussolotti}
\affiliation{Graduate School of Advanced Integration Science, Chiba University, 1-33 Yayoi-cho, Inage-ku, Chiba 263-8522, Japan}

\author{Kouki Akaike}
\affiliation{Photoelectric conversion research team, Advanced science institute, RIKEN, 2-1 Hirosawa, Wako-shi, Saitama 351-0198, Japan}

\author{Yoshihiro Kubozono}
\affiliation{Research Laboratory for Surface Science \& Research Center of New Functional Materials for Energy Production, Storage and Transport, Okayama University, Okayama 700-8530, Japan}

\author{Hideo Aoki}
\affiliation{Department of Physics, University of Tokyo, Hongo, Tokyo 113-0033, Japan}

\author{Taichi Kosugi}
\affiliation{Nanosystem Research Institute ``RICS'', AIST, Umezono, Tsukuba 305-8568, Japan}

\author{Satoshi Kera}
\author{Nobuo Ueno}\email{uenon@faculty.chiba-u.jp}
\affiliation{Graduate School of Advanced Integration Science, Chiba University, 1-33 Yayoi-cho, Inage-ku, Chiba 263-8522, Japan}

\date{\today}

\begin{abstract}
We have experimentally revealed the band structure and the surface Brillouin zone of insulating picene single crystals (SCs), the mother organic system for a recently discovered aromatic superconductor, with ultraviolet photoelectron spectroscopy (UPS) and low-energy electron diffraction with laser for photoconduction. A hole effective mass of 2.24\,$m_{0}$ and the hole mobility $\mu_{\textrm{h}}$\,$\geq$\,9.0\,cm$^{2}$/Vs (298\,K) were deduced in $\Gamma$-$Y$ direction. We have further shown that some picene SCs did not show charging during UPS even without the laser, which indicates that pristine UPS works for high-quality organic SCs.

\end{abstract}

\maketitle

Organic single crystals (SCs) have recently attracted considerable attention both for fundamental research and device applications because of some properties not found in inorganics and outstanding functionalities in optoelectronic devices. Organic SCs show highly anisotropic physical properties; for planar conjugated molecules, for instance, the highest charge-carrier mobility usually occurs in the $\pi$-stacking direction along which the electronic coupling is strongest. Studies of the electronic band structure of organic SCs are thus critical in unraveling their charge transport properties. In contrast to vacuum-sublimated thin films, however, organic SCs commonly exhibit serious charging upon ionizing UV irradiation or electron bombardment, which hinders ultraviolet photoelectron spectroscopy (UPS) or low-energy electron diffraction (LEED) investigations to probe the band structure and the crystal orientation. It was demonstrated for UPS that the charging can be overcome by concomitant illumination by UV and laser light \cite{Sato1985JCP,Zimmermann2000mclc,Vollmer2005EPJE}, and recently some of the present authors and coworkers succeeded in measuring the band dispersion of rubrene SCs at room temperature by angle-resolved UPS (ARUPS) \cite{Machida2010PRL}. However, to our knowledge, no LEED study of an organic SC exists so far, although LEED is a straightforward approach to access the surface Brillouin zone and thus the orientation of a SC.

Recently, picene [C$_{22}$H$_{14}$, Fig.\,1(a)] has attracted much attention because the solid picene was found to be the first aromatic superconductor upon potassium-doping with a T$_{\textrm{c}}$ (18\,K) that is very high as an organic superconductor \cite{Mitsuhashi2010N,Kubozono2011pccp}, and also because high hole mobilities ($\sim$5\,cm$^{2}$/Vs) were found even in thin film transistors\cite{Okamoto2008,Kawasaki2009}. Moreover, picene has higher chemical stability than its isomer, pentacene, as indicated by a larger optical gap [3.11\,eV; Fig.\,1(a)] than that of pentacene (1.85\,eV) \cite{Salzmann2007PRB}. So it is highly desirable to probe the electronic structure of picene SCs. In the present work we have succeeded in measuring LEED of undoped picene SCs by overcoming the sample charging with laser illumination. This has enabled us to determine accurately the orientation of the SC and thus to measure ARUPS along defined crystallographic directions. Surprisingly, some picene SCs (2 out of 10 under study) had no charging problem during UPS measurements even without laser, which indicates that UPS can directly measure the electronic structure of insulating organic SCs if their quality is very high with ultralow density of charge-trapping states.

\begin{figure}
\centering
\includegraphics[scale=1]{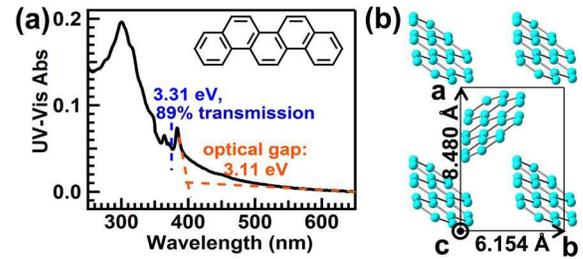}
\caption{\label{molecules}(a) Chemical structure of picene and the UV$\-$Vis absorption spectrum of a picene thin film ($\thicksim$20\,nm) on quartz. The blue dashed line indicates the photon energy of the laser used in the experiment; (b) Single crystal structure projected onto the $ab$ plane \cite{De1985acc}.}
\end{figure}

Plate-like picene SCs ($\sim$2\,$\times$\,2\,$\times$\,0.15\,mm$^{3}$) with (001) orientation [Fig.\,1(b)] were grown by physical vapor deposition in a purified argon stream as described elsewhere \cite{Mitsuhashi2010N}. The SCs were bound on Cu plate substrates with silver paste (DOTITE, D-550; FUJIKURAKASEI Co.) in air [see Fig.\,S1(a) of the Supporting Information] , and then transferred to the ultrahigh vacuum chamber ($\sim$8\,$\times$\,10$^{-8}$\,Pa) for LEED, UPS and metastable atom electron spectroscopy (MAES) measurements at room temperature. LEED measurements were carried out with a microchannel plate (OCI Vacuum Microengineering) and an assistant laser (3.31\,eV, incident angle $\alpha_{\textrm{L}}$\,$\approx$\,80\,$^\circ$). For calibration of the lattice constants an Cu(111) crystal was used. ARUPS and MAES were performed in a custom-built apparatus \cite{Duhm2012am} with an electron energy analyzer (Scienta R3000) using mainly p-polarized He I radiation (21.22\,eV) and metastable He* atoms (2$^{3}$S, 19.82\,eV) as the excitation sources, respectively. Due to the small size of the SC a spatially resolved lens mode with an emission acceptance angle of ($\pm$15$^\circ$, $\pm$5$^\circ$) [as shown in Fig.\,S1(b) of the Supporting Information] was used for the ARUPS and MAES in order to selectively measure the electrons from the SC. High resolution UPS (HRUPS) spectra were measured with an electron energy analyzer (MBS A-1) and a monochromatic He I¦Á (21.22\,eV) light source \cite{Sueyoshi2009APL}. For ARUPS, the incident angle ($\alpha$) of He I was fixed to 65$^\circ$, the emission angle ($\theta$) varied between 0$^\circ$ and 42$^\circ$, and the incident angle of the laser was $\alpha_{\textrm{L}}$\,=\,45$^\circ$\,$-$\,$\theta$. For MAES $\alpha$\,=\,45$^\circ$, $\alpha_{\textrm{L}}$\,=\,45$^\circ$ and $\theta$\,=\,0$^\circ$. HRUPS was measured at $\alpha$\,=\,45$^\circ$, $\theta$\,=\,0$^\circ$. The LEED, ARUPS, MAES and HRUPS experimental geometries are shown in Fig.\,S1(b) of the Supporting Information. The energy resolution was set to 80\,meV for ARUPS and 30\,meV for HRUPS. The position of the Fermi level ($E_{\textrm{F}}$) was determined from an Ag(111) crystal. The UV-Vis absorption on a 20-nm picene film evaporated on quartz was measured using a JASCO U-570 spectrophotometer [Fig.\,1(a)]. Fitting of UPS spectra was done using WinSpec developed at Namur University, Belgium.


No clear LEED pattern could be measured without using laser illumination, since the crystal exhibited strong charging [Fig.\,2(a)] under electron bombardment in an electron energy range from 5 to 50\,eV. However, the charging was found to be overcome by concomitant laser irradiation accompanied by bright pale blue luminescence of the sample [Fig.\,S1(a) of the Supporting Information], where clear diffraction spots [with light background which may be due to the luminescent of the SC, Fig.\,2(b)] could be measured at an electron energy of 36\,eV with a sample current of $\sim$55\,pA. By switching the laser on and off, the charging could be simultaneously (on a time scale of human eye response) switched off and on. For resolving the charging a laser power of 2.8\,mW was sufficient, and further increasing the laser power did not change the LEED pattern. The lattice parameters of SC \#1 surface were evaluated as $a$\,=\,8.6\,{\AA}, $b$\,=\,6.3\,{\AA}, $\gamma$\,=\,90$^\circ$, which correspond well with the reported lattice constants ($a$\,=\,8.480\,{\AA}, $b$\,=\,6.154\,{\AA}, $c$\,=\,13.515\,{\AA}, $\alpha$\,=\,$\gamma$\,=\,90$^\circ$, and $\beta$\,=\,90.46$^\circ$) in the $ab$ plane \cite{De1985acc}, suggesting no notable reconstruction occurs at the SC surface. The LEED measurement allowed an azimuthal arrangement of the SC and thus ARUPS measurements along well defined crystallographic directions.

\begin{figure}
\centering
\includegraphics[scale=1]{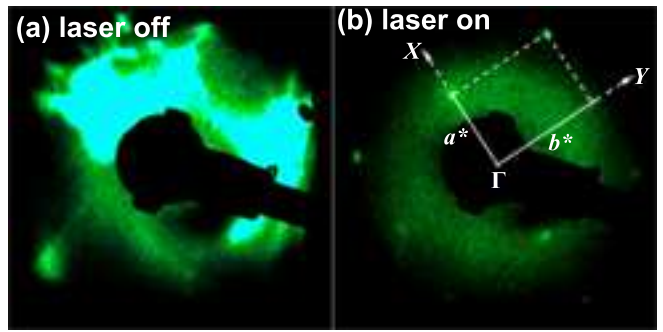}
\caption{\label{LEED}LEED pattern of picene SC \#1 with laser off (a) and on (b), both with an electron energy of 36\,eV.}
\end{figure}

During UPS measurements, 2 out of 10 samples (e.g.\,\#2) did not show a charging effect even without laser illumination, while for the other 8 samples the charging could be totally (e.g.\,\#1) or partly suppressed by the laser. This behavior may be caused by different crystal qualities and/or probable interface contacting differences. The normal emission (i.e.\,$\Gamma$-point) ARUPS and MAES spectra of \#1 (with laser) and the HRUPS of \#2 (without laser) are displayed in Fig.\,3. Since metastable He* atoms do not penetrate the sample surface, MAES just probes the occupied molecular orbitals of the outermost surface \cite{Harada1997CR}. The well featured MAES spectra indicate that, although no in-situ cleaning procedure was applied to the SCs, the surface was not contaminated notably with ambient moieties.

The ARUPS and MAES features of \#1 show excellent consistency, and agree well with the reported UPS spectrum of an evaporated thin film on Au \cite{Roth2010njp}. Without laser SC \#1 got charged under HeI light [Fig.\,S2 of the Supporting Information] or He* bombardment, while with a laser power of 9.8\,mW the charging could be totally resolved. For the charging-free SC (\#2) the laser induced only a higher secondary-electrons background (probably due to electron-exciton inelastic scattering) and a slight energy level shift ($\sim$60\,meV) to higher binding energy ($E_{\textrm{B}}$) by the photovoltaic effect \cite{Koch2003PRB} [Fig.\,S2 of the Supporting Information]. Here, the laser illumination for the charging SCs (e.g. \#1) induces a decharging effect which sharpens the UPS features significantly and shifts the spectra to the lower $E_{\textrm{B}}$, whereas for the charging-free SCs (e.g. \#2) it induces a slight photovoltaic effect which shifts the UPS spectra to the higher $E_{\textrm{B}}$. The whole valence band of \#2 (without laser) was observed at lower $E_{\textrm{B}}$ by 0.17\,eV than that of \#1 (Fig.\,3 and Fig.\,S2 of the Supporting Information). For all measured SCs the profile of the UPS spectra (at the same measurement geometry) was almost the same, whereas the absolute $E_{\textrm{B}}$ shifted within a range of 0.3\,eV. This shift may be caused by the different crystal quality. The SCs with higher quality have a lower density of gap states (charge trapping states), which may shift the $E_{\textrm{F}}$ closer to the HOMO \cite{Sueyoshi2009APL,Mao2011OE,Hosokai2011}. SC \#2 did not show any clear density of occupied states in the energy gap [Fig.\,S2(c) of the Supporting Information], although it was measured with an experimental set-up which is designed to measure tiny gap states \cite{Sueyoshi2009APL}, suggesting an extremely low density of defects in the crystal.
%
\begin{figure}
\centering
\includegraphics[scale=1]{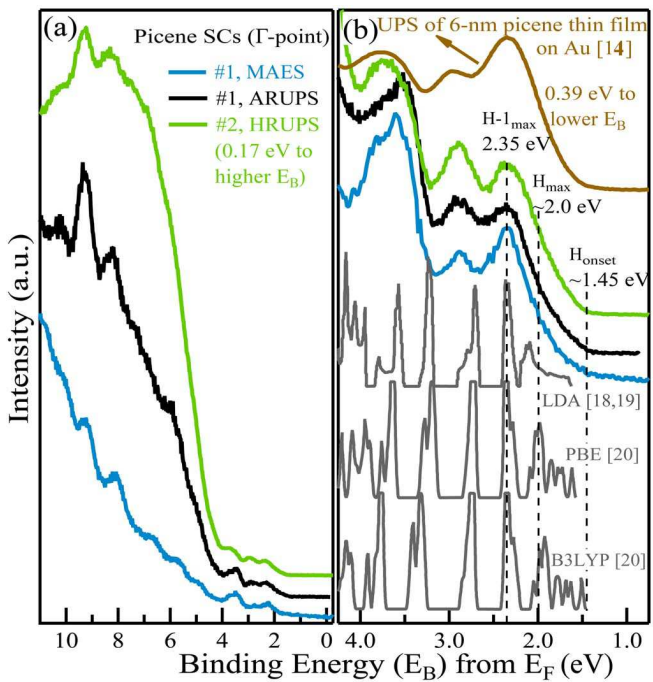}
\caption{\label{UPS and MAES}(a) Normal emission UPS and MAES spectra of picene SC \#1 with laser and HRUPS spectrum of \#2 without laser; (b) A close-up of the low $E_{\textrm{B}}$ region of (a), the UPS spectrum of a picene thin film (6\,nm) on Au \cite{Roth2010njp} and the first-principles density of states (DOS) of single crystalline picene calculated with three different methods: local density approximation (LDA; present result, see Ref.\cite{Kosugi2011prb,Kosugi2009jpsj} for theoretical method), Perdew-Burker-Ernzerhof (PBE) and B3LYP \cite{Giovannetti2011prb}. The calculated DOS was shifted in binding energy to match the experimental HOMO-1 derived peak.}
\end{figure}

The first-principles band structure of single crystalline picene, first obtained in \cite{Kosugi2011prb} and shown here for three different implementations \cite{Kosugi2009jpsj,Kosugi2011prb,Giovannetti2011prb} of the exchange functional etc, is also displayed in Fig.\,3(b). Apart from slight differences between the three methods, they agree fairly well with the experimental result for the $E_{\textrm{B}}$\,$\lesssim$\,3\,eV.

As for the band dispersion, the theoretical result in Fig.\,4(c) shows a relatively large bandwidth ($W$) of the two HOMO derived bands: $\sim$0.24\,eV along $\Gamma$-$X$ and $\sim$0.51\,eV along $\Gamma$-$Y$ direction for the upper HOMO band and $\sim$0.24\,eV along $\Gamma$-$X$ and $\sim$0.06\,eV along $\Gamma$-$Y$ direction for the lower HOMO band \cite{Kubozono2011pccp,Kosugi2009jpsj,Kosugi2011prb}. On the other hand, ARUPS along $\Gamma$-$Y$ direction has been studied and the results are shown in Fig.\,4(a). The peak centered at ~2.35\,eV was assigned to be originated from the HOMO-1, while the HOMO derived peak gets nearly masked in the slope of the HOMO-1 derived peak [see also Fig. 3\,(b)]. Since the angular acceptance is $\pm$15$^\circ$ along $\Gamma$-$X$ direction, the ARUPS spectra along $\Gamma$-$Y$ simultaneously cover the $\Gamma$-$X$ direction. Moreover the angular acceptance is $\pm$5$^\circ$ along $\Gamma$-$Y$ direction, and this additionally blurs the band dispersion measurement in $\Gamma$-$Y$ direction. The ARUPS top feature (HOMO and HOMO-1 derived peaks) was thus deconvoluted with four Voigt (90$\%$ Gaussian + 10$\%$ Lorentzian) functions: a constant component $\textrm{H}_{\textrm{X}}$ for the HOMO contribution in $\Gamma$-$X$ direction, a dispersed component $\textrm{H}_{\textrm{Y}}^{\textrm{u}}$ for the upper HOMO band in $\Gamma$-$Y$ direction, a constant component $\textrm{H}_{\textrm{Y}}^{\textrm{l}}$ for the lower HOMO band in $\Gamma$-$Y$ direction, and H-1 for HOMO-1 [see Fig.\,4(a)]. The HOMO-2 derived peak was deconvoluted also with a Voigt function [H-2 in Fig.\,4(a)]. The dispersions of the HOMO components $\textrm{H}_{\textrm{X}}$, $\textrm{H}_{\textrm{Y}}^{\textrm{u}}$ and $\textrm{H}_{\textrm{Y}}^{\textrm{l}}$ are shown in Fig.\,4(b) as a function of $k_{\parallel}$, the surface parallel component of the electron wave vector, where the error of $k_{\parallel}$ is estimated to be $\Delta$$k_{\parallel}$\,$\leq$\,$\pm$0.17\,{\AA}$^{-1}$ \cite{Ueno2008PSS}. The fitting shows that $\textrm{H}_{\textrm{Y}}^{\textrm{u}}$ has a dispersion width of $\sim$0.18\,eV [Fig.\,4(b)]. In addition, we did not observe clear signature of the HOMO-band dispersion along $\Gamma$-$X$. The experimental HOMO-band widths along $\Gamma$-$Y$ and $\Gamma$-$X$ are smaller than the calculated result, which is probably caused by experimental restrictions as well as the error in deconvolution. However, our results show that (i) there is unambiguously a HOMO dispersion along $\Gamma$-$Y$, and (ii) the overall shape of the dispersion agrees with the calculation.
%
\begin{figure}
\centering
\includegraphics[scale=1]{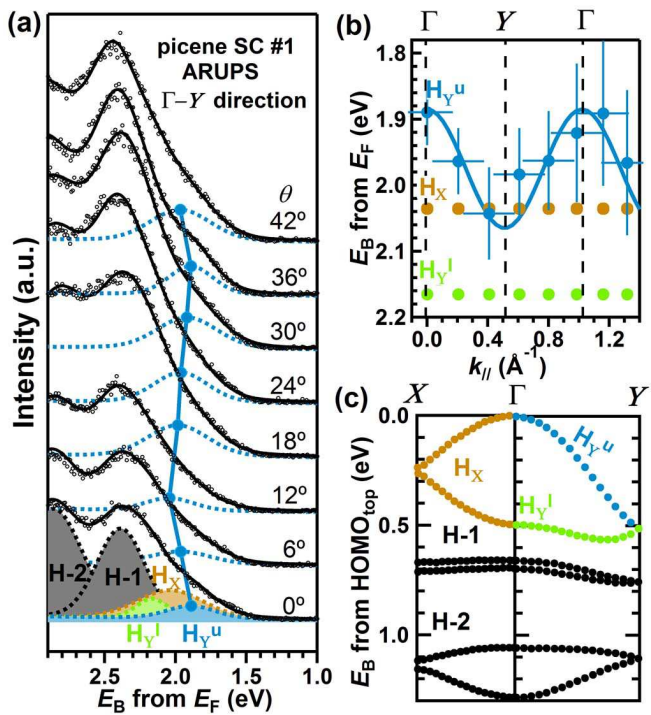}
\caption{\label{band dispersion}(a) ARUPS of SC \#1 along $\Gamma$-$Y$ direction with deconvoluted $\textrm{H}_{\textrm{X}}$ (HOMO component in $\Gamma$-$X$ direction, brown dotted curve with shade), $\textrm{H}_{\textrm{Y}}^{\textrm{u}}$ (the upper HOMO band component in $\Gamma$-$Y$ direction, blue dotted curve with shade), $\textrm{H}_{\textrm{Y}}^{\textrm{l}}$ (the lower HOMO band component in $\Gamma$-$Y$ direction, green dotted curve with shade), H-1 and H-2 (HOMO-1 and HOMO-2, black dotted curves with shade) Voigt functions. The $\textrm{H}_{\textrm{X}}$, $\textrm{H}_{\textrm{Y}}^{\textrm{l}}$, H-1, H-2, and the shade part of $\textrm{H}_{\textrm{Y}}^{\textrm{l}}$ of $\theta$\,=\,6\,$\sim$\,42$^\circ$ and are omitted for clarity; (b) The dispersion of $\textrm{H}_{\textrm{X}}$, $\textrm{H}_{\textrm{Y}}^{\textrm{u}}$ and $\textrm{H}_{\textrm{Y}}^{\textrm{l}}$ with respect to the parallel component of electron wave vector $k_{\parallel}$. The solid curve is a fit to the tight-binding model; (c) A part of the calculated valence band structure of picene SC along $\Gamma$-$Y$ and $\Gamma$-$X$ \cite{Kubozono2011pccp,Kosugi2009jpsj,Kosugi2011prb}.}
\end{figure}

By assuming a tight-binding model, the dispersion of the upper HOMO band in $\Gamma$-$Y$ direction $E_{\textrm{Y}}^{\textrm{u}}$ can be expressed as $E_{\textrm{Y}}^{\textrm{u}}$($k_{\parallel}$)\,=\,$E_{0}^{\textrm{u}}$\,$-$\,2$t$$\cos$($b$$k_{\parallel}$), where $E_{0}^{\textrm{u}}$ is the band center of the upper HOMO band, $t$ the transfer integral, and $b$ the lattice constant. Fitting this dispersion with the experimental results [Fig.\,4(c)] gives $E_{0}^{\textrm{u}}$\,=\,1.975\,eV and $t$\,=\,45\,meV. In the tight-binding approximation, the effective mass of the hole $m_{\textrm{h}}$$^{*}$ can be given by $m_{\textrm{h}}$$^{*}$\,=\,$\hbar^{2}/2tb^{2}$, while our results gives $m_{\textrm{h}}$$^{*}$\,=\,2.24\,$m_{0}$ ($m_{0}$: the free electron mass). In a broad band model ($W$\,$>$\,$k_{\textrm{B}}$$T$), the hole mobility $\mu_{\textrm{h}}$ can be estimated as $\mu_{\textrm{h}}$\,$\geq$\,20\,($m_{0}/m^{*}$)\,$\times$\,(300/$T$) \cite{Froehlich1959ppsl}, so that the lower limit of $\mu_{\textrm{h}}$ is 9.0\,cm$^{2}$/Vs at 298\,K. This value is comparable to the largest $\mu_{\textrm{h}}$ ($\sim$5\,cm$^{2}$/Vs) of reported picene thin film transistors \cite{Okamoto2008,Kawasaki2009}, and suggests that the hole mobility in the thin films is dominated by the band transport \cite{Okamoto2008,Kawasaki2009,Roth2010njp,Wang2011jacs}.

Besides this insight in the band structure of picene, the most astonishing result is that the negative (during LEED) and positive (during UPS and MAES) sample charging can be overcome by the same approach, i.e., laser illumination with the energy of 3.31\,eV, which is larger than the optical gap [3.11\,eV, Fig.\,1(a)] but smaller than the transport band gap of picene (assuming a typical exciton binding energy in the range of 0.4$\sim$1.4\,eV \cite{Knupfer2003apa,Djurovich2009oe}). The 20-nm picene thin film can transmit 89$\%$ of light with the photon energy of 3.31\,eV [Fig.\,1(a)]. This implies that for SCs with a thickness of $\sim$0.15\,mm, the transmitted laser is negligible and the photoconductivity might be related to migration of molecular excitons to the back surface of the SC followed by injection of the counter charges to the SC. Moreover, the accordance of the UPS spectra measured with and without laser for the charging-free samples demonstrates that the laser only leads to a slight photovoltaic effect.

In conclusion, with laser illumination, the surface Brillouin zone, the electronic structure and the HOMO band dispersion along the $\pi$-stacking direction ($\Gamma$-$Y$) were determined for picene single crystals by means of LEED, UPS and MAES. The laser illumination can overcome the sample charging, and does not influence the valence band structure (negligible photovoltaic effect). The LEED results elucidate the surface Brillouin zone that corresponds well with the bulk Brillouin zone of the $ab$ plane, and probably no notable surface reconstruction occurs. Our results highlight that upon laser illumination combining LEED and ARUPS is an available way to obtaining the band structure along certain crystallographic direction of organic single crystals, and this technique could also be applied to other insulators. Moreover, the electronic structure of high quality organic crystals can be measured by UPS without using photoconduction. This offers an important guideline to understand the true origin of the charging phenomena and to realize electron spectroscopy study of other insulating SCs. As for the band dispersion on which a comparison with first-principles results was done here, if we consider the electron correlation effects, they not only renormalize the effective mass, but also change the shape of the DOS \footnote{Such an effect is studied e.g. for the iron-based superconductor in W. Malaeb, T. Yoshida, T. Kataoka, A. Fujimori, M. Kubota, K. Ono, H. Usui, K. Kuroki, R. Arita, H. Aoki, Y. Kamihara, M. Hirano, and H. Hosono, J. Phys. Soc. Jpn. \textbf{77}, 093714 (2008).}, which will also merit further investigations.

The authors thank Dr.\,Martin Oehzelt, Dr.\,Antje Vollmer (both at Helmholtz Zentrum Berlin) and Dr.\,Alexander Hinderhofer (Chiba University) for fruitful discussions, Xu Lin and Prof.\,Shiki Yagai (both Chiba University) for the optical absorption measurement, and Dr.\,Yuli Huang and Weining Han (both Chiba University) for experimental support. HA and TK acknowledge discussions on the first principles band structure with Prof.\,Ryotaro Arita (University of Tokyo) and Drs. Takashi Miyake, Shoji Ishibashi (both at AIST). This work was financial supported partly by a Grant-in-Aid for Scientific Research (A) (Nos. 20245039, 19051016 and 22104010), Next Generation Supercomputer Project, Nanoscience Program from MEXT, Japan, and the Global-COE program of MEXT (G3: Advanced School for Organic Electronics, Chiba University). HA is supported by an EU-Japan ``LEMSUPER'' project. SK gratefully acknowledges support from a Grant-in-Aid for Young Scientists (A) (20685014).

\end{document}